\documentclass[prb,showpacs,twocolumn]{revtex4-1}
\usepackage{graphicx}
\usepackage{dcolumn}
\usepackage{bm}
\usepackage{hyperref}

\begin{document}
\title{Charge order at magnetite Fe$_3$O$_4$(001): surface and Verwey phase transitions}

\author{I.~Bernal-Villamil, S.~Gallego} 
\email{sgallego@icmm.csic.es}
\affiliation{
Instituto de Ciencia de Materiales de Madrid, Consejo Superior de
Investigaciones Cient{\'{\i}}ficas, Cantoblanco, 28049 Madrid, Spain}

\date{\today}

\begin{abstract}
At ambient conditions, the Fe$_3$O$_4$(001) surface shows a $(\sqrt{2}\times\sqrt{2})R45^o$ reconstruction 
that has been proposed as the surface analog of the bulk phase below the Verwey transition temperature, T$_V$.
The reconstruction disappears at a high temperature, T$_S$, through a second order transition.
We calculate the temperature evolution of the surface electronic structure based on a reduced 
bulk unit cell of $P2/m$ symmetry that contains the main features of the bulk charge distribution.
We demonstrate that the insulating surface gap arises from the large demand of charge of the surface O,
at difference with that of the bulk. Furthermore, it is coupled to a significant restructuration 
that inhibits the formation of trimerons at the surface.
An alternative bipolaronic charge distribution emerges below T$_S$, introducing
a competition between surface and bulk charge orders below T$_V$.
\end{abstract}

\maketitle

\section{Introduction}

Magnetite (Fe$_3$O$_4$) is the oldest known magnet and a 
fascinating material both for understanding the fundamental physics that emerge from electronic correlations,
and for novel technologies related to oxide electronics \cite{Orozco-1999,Wu-2013,Liu-2013}.
At ambient conditions, it crystallizes in the inverse spinel structure with cubic $Fd\overline{3}m$ symmetry.
The O atoms form a fcc lattice, with Fe$_A$ atoms in tetrahedral sites acting with a nominal $+3$ valence, and 
Fe$_B$ cations with $+2.5$ valence in octahedral positions.
The Fe$_A$ and Fe$_B$ sublattices are antiferromagnetically coupled, 
and the minority spin $t_{2g}$ states of the Fe$_B$ atoms cross the Fermi level,
leading to a half-metallic ferrimagnet with high 
magnetic moment, 4 $\mu_B$ per formula unit (f.u.).

At a critical temperature T$_V \sim 120$ K, magnetite undergoes the first-order Verwey transition (VT),
that manifests in a structural modification to a monoclinic symmetry
accompanied by a drop of the conductivity of 2-3 orders of magnitude \cite{Verwey}. The decrease of the conductivity
is due to a freezing of the electron hopping between different octahedral Fe sites, causing
a charge disproportionation that results in two types of Fe$_B$ atoms acting with a slightly enhanced (Fe$^{3+}$)
or reduced (Fe$^{2+}$) valence. The distribution of the different Fe$_B$ atoms at the unit cell configures 
the charge order (CO), intimately linked to the orbital order \cite{Leonov-2004,Huang-2006},
and determines the full monoclinic $Cc$ symmetry \cite{Iizumi-1982,Jeng-2006,Yamauchi-2009,Garcia-2011,Attfield-nature}.
Decades of efforts have been devoted to the understanding of the VT \cite{Walz-2002,Garcia-2004}, 
and yet some puzzling fundamental aspects remain
unanswered, such as the structural or electronic origin of the transition, or the extent of the
short range order above T$_V$. The present consensus is that 
the phase transition is governed by electron-phonon couplings in the presence of strong electronic correlations
\cite{Piekarz-2013}. 
A local perturbation of the extended CO has been recently identified in the form of trimerons: linear chains of three adjacent 
Fe$_B$ cations dominantly formed by a central Fe$^{2+}$ and two Fe$^{3+}$, with a significant reduction of the interatomic Fe-Fe distances
and a polaronic distribution of shared charge \cite{Attfield-nature}.
Trimerons reveal as the essential short-range unit in the electronic phase transitions of magnetite \cite{Piekarz-2014}. 
Furthermore, laser pump-probe experiments have created a non-equilibrium version of the VT by introducing holes in the trimeron lattice \cite{Jong-2013}.
This invokes the possibility to obtain analogs of the VT under sizes much lower than those required by a full $Cc$ cell,
and in fact, as a first result of this work, we will demonstrate the ability of trimerons to emerge in a reduced unit cell of $P2/m$ symmetry.

One of the handicaps for the exploitation of the VT in novel technologies is the low value of T$_V$, well below 
room temperature (RT).
The measurement at the Fe$_3$O$_4$(001) surface of an insulating gap 
at RT \cite{Jordan-2006} and the further prediction that its existence
was accompanied by a subsurface CO similar to the bulk one \cite{Lodziana-2007}, caused thus great excitement.
Fe$_3$O$_4$(001) presents a $(\sqrt{2}\times\sqrt{2})R45^o$ reconstruction 
which corresponds to a bulk truncation at an Fe$_B$-O 
plane \cite{Stanka-2000,Shvets-2004,Novotny-2013}. Its origin has been assigned to a Jahn-Teller distortion causing
a wavelike displacement of the Fe$_B$ surface atoms along $<110>$ directions \cite{Pentcheva-2005,Pentcheva-2008}.
Seemingly this RT
reconstruction is not significantly altered across the VT \cite{Pentcheva-2008,JdF-roof}. However, the evolution in
depth of the subsurface CO has not been investigated, setting forth interrogants about the formation of surface trimerons
and the relation between the surface and bulk COs below T$_V$. 
A recent study proves that the bulk low temperature phase (LTP) manifests at the surface in distinct structural features 
than the reconstruction \cite{JdF-roof}.
Furthermore, the $(\sqrt{2}\times\sqrt{2})R45^o$ symmetry is lost in favor of a $(1 \times 1)$ structure
at a temperature T$_S \sim 720$ K through a second-order transition involving loss of long-range CO
\cite{Norm-2013}.
These results suggest the existence of fundamental differences between the surface and bulk insulating phases.

In this work we provide firm proof of this fact, calculating the evolution with temperature of the electronic structure 
of the Fe$_3$O$_4$(001) surface. 
Our results evidence that the surface insulating state 
originates from the combination of large O electron affinity and loss of O bonds,
and remains unaltered across the bulk and surface transitions.
This has an impact for the disappearance of trimerons close to the surface,
replaced by bipolaronic structures below T$_S$. As a consequence,
a competition between the local bulk and surface COs emerges below T$_V$, that manifests in
modulations of the surface CO arising both from bulk trimerons and from the surface 
reconstruction.

\section{Theoretical method}

We have performed first principles calculations of both bulk magnetite and its (001) termination,
based on the density functional theory including correlation effects. 
We employ a plane wave basis set \cite{vasp1} and the projector augmented waves (PAW) method to describe the
core electrons \cite{paw}, with an energy cutoff of 400 eV and a Monhorst-Pack sampling of the Brillouin zone (BZ) 
of $(7 \times 7 \times 5)$ for the bulk and up to $(6 \times 6 \times 2)$ for the surface slabs,
that guarantee convergence in the total energy better than 0.1 meV/f.u.
We use the exchange-correlation functional parametrization of Perdew-Burke-Erzenhof (PBE), adding
an effective on-site Coulomb repulsion term U=4 eV \cite{dudarev}. This choice of U is based on the recovery of
an equilibrium value of the cubic lattice parameter $a=8.4$ $\text{\AA}$ in excellent agreement with experiments, and
the adequate description of the Verwey transition in terms of charge disproportionation (0.27 $e$) and electronic
band gap (0.2 eV) when reducing the symmetry from the cubic $Fd\overline 3m$.

Our description of bulk magnetite is based on a $P2/m$ unit cell formed by 28 atoms.
We have determined the equilibrium structures above and below T$_V$ starting from the ideal cubic lattice and allowing relaxation
of the lattice vectors and atomic positions, with no symmetry constraints for the low temperature phase (LTP).
Even at the high temperature phase (HTP) there exists a noticeable distortion of the O sublattice, that introduces a slight tetragonal deformation 
of the unit cell with a small reduction of the total volume of 3 $\text{\AA}^3$. 
At the LTP, relaxation of the lattice vectors leads to an orthorhombic symmetry, but again the distortion of the unit cell is small,
with a similar reduction of the total volume.

To model the Fe$_3$O$_4$(001) surface, we have used slabs of different thicknesses, containing from 8 to 16
atomic planes, supported on a Au substrate and including a vacuum region of at least 12 $\text{\AA}$ that avoids interaction
between opposite slab surfaces. The choice of the substrate has been performed to
minimize interface effects and to confine them to the interface layer. In all cases we have employed
$(\sqrt{2}\times\sqrt{2})R45^o$ two-dimensional unit cells, starting our calculations either from
the $(1 \times 1)$ termination or from the Jahn-Teller induced wavelike pattern,
and allowing to relax the atomic positions of the 3 outermost surface layers until the forces on all atoms are
below 0.01 eV/$\text{\AA}$. We have done this for slabs constructed both from the HTP and the LTP bulk structures.
The slabs of 12 planes provide the minimum thickness to recover the bulk structure at the inner layers below T$_V$
including the distribution of trimerons, and all the results presented here correspond to this configuration.
We have also modelled thicker unsupported symmetric slabs of 16 planes to check the independence of our
conclusions on the slab configuration, particularly concerning the penetration of surface effects.

\section{Bulk $\text{Fe}_3\text{O}_4$}

\begin{figure}[htbp]
\begin{center}
\includegraphics[width=\columnwidth,clip]{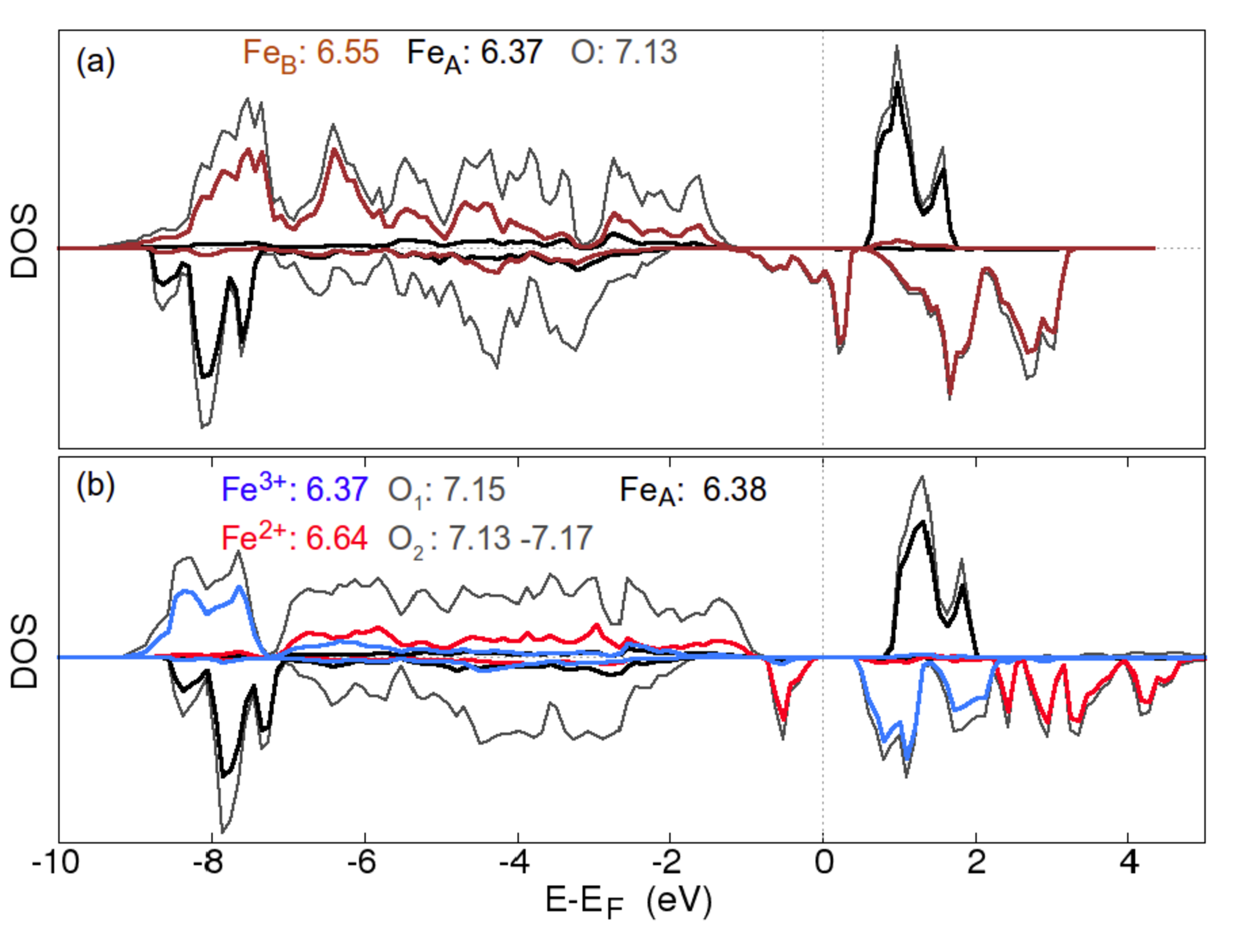}
\caption{
(Color online)
Total DOS of bulk Fe$_3$O$_4$ at the (a) HTP and (b) LTP of bulk magnetite, showing the projections on the
Fe$_A$ (thick black) and inequivalent Fe$_B$ (red/blue) sites.
Positive (negative) DOS values correspond to majority (minority) spin projections.}
\label{fbulk}
\end{center}
\end{figure}

Figure \ref{fbulk} and Table \ref{tabla-dist} summarize our results
for the density of states (DOS), the Bader charges (Q$_B$) and the interatomic distances at 
both the HTP and the LTP of bulk magnetite. The energy barrier between both phases is 170 meV/f.u.
The higher symmetry of the HTP reflects in 
the existence of only one type of O and Fe$_B$ sites with a bond length of 2.06 $\text{\AA}$, and
in the uniform value of the Fe$_B$-Fe$_B$ interatomic distance (d$^{FF}=2.96\text{\AA}$).

\begin{figure}[htbp]
\begin{center}
\includegraphics[width=.65\columnwidth,clip]{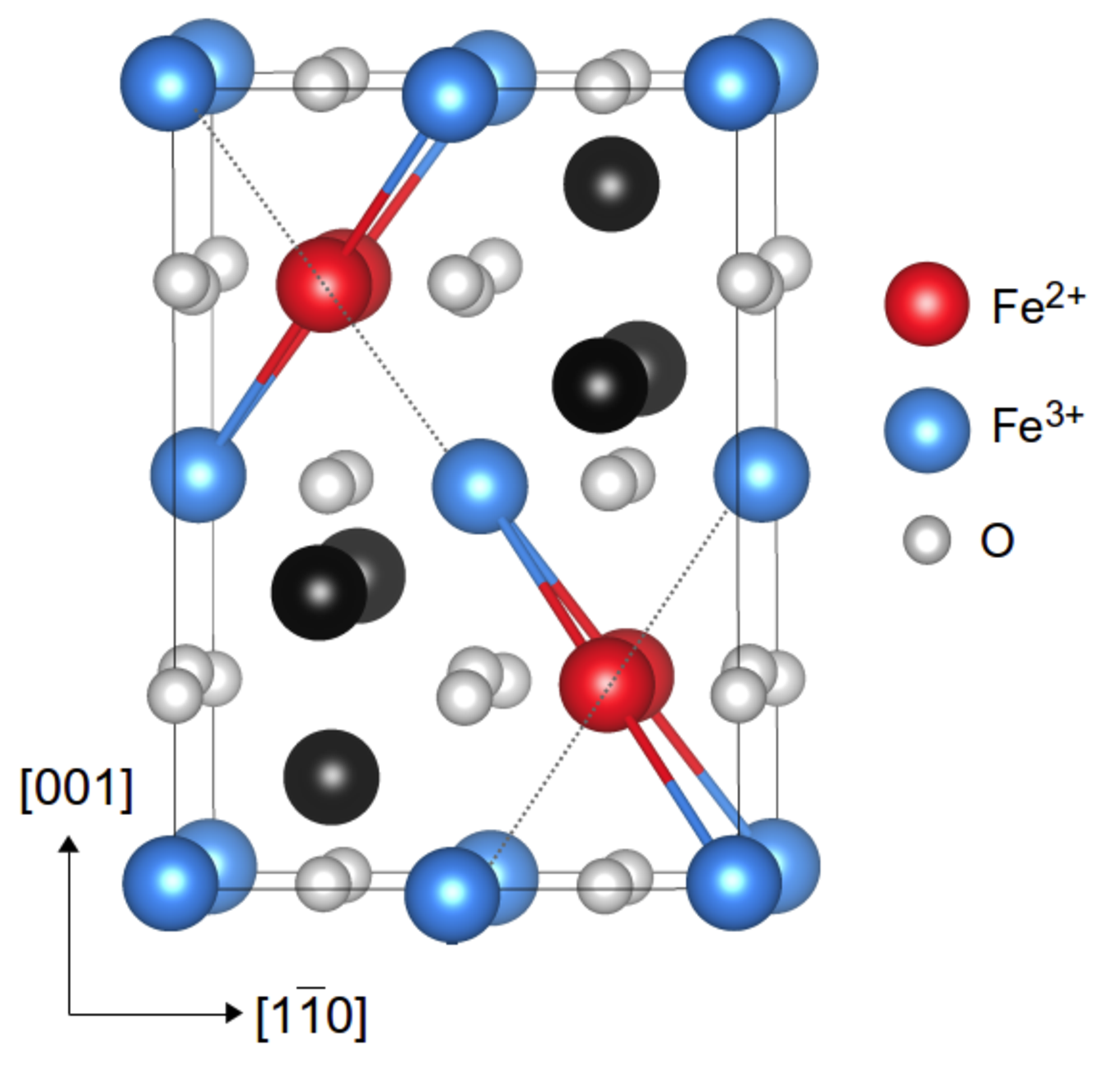}
\caption{Distribution of trimerons at the $P2/m$ unit cell of the LTP of bulk Fe$_3$O$_4$.}
\label{fbulk-trims}
\end{center}
\end{figure}

Below T$_V$, while Fe$_A$ remains essentially unaffected by the transition, 
a charge disproportionation of 0.27 $e$ appears in the Fe$_B$ sublattice, 
opening a band gap of 0.2 eV.
Within our reduced $P2/m$ cell, 
the Fe$^{2+}$ and Fe$^{3+}$ ions alternate along the [001] direction, as evidenced in figure \ref{fbulk-trims}.
Different values of the d$^{FF}$ can be found depending on the Fe valence.
This is accompanied by a noticeable dispersion of the Fe$_B$-O bond lengths, 
with larger average values for Fe$^{2+}$ (2.08$\text{\AA}$) than for Fe$^{3+}$ (2.03$\text{\AA}$).
The result is a non-uniform distribution of charge and magnetic moments that leads to slightly different
O atoms at the Fe$^{3+}$ (O$_1$) and Fe$^{2+}$ (O$_2$) planes, 
as reflected in the dispersion of the Q$_B$ values.
However, the same net magnetization of 4 $\mu_B$/f.u. is obtained above and below T$_V$.
\begin{table}[hbtp]
\begin{center}
\caption{Mean Fe-O bond-lengths (d(Fe-O)) and values of the interatomic distances between first Fe$_B$ neighbors 
(d$^{FF}$) at the HTP and LTP of bulk Fe$_3$O$_4$. Units are $\text{\AA}$.
 \label{tabla-dist}}
\renewcommand{\arraystretch}{1.5} 
\renewcommand{\tabcolsep}{0.4pc}
\begin{tabular}{c c c c }
\hline
\hline
d(Fe-O)	&Fe$^{2+}$-O		&Fe$^{3+}$-O		&Fe$_A$-O  \\ \hline
HTP					&2.06			&\textendash\ 	&1.89 \\
LTP					&2.03			&2.08			&1.89 \\ 
\hline
\hline
d$^{FF}$	&Fe$^{2+}$-Fe$^{2+}$	&Fe$^{2+}$-Fe$^{3+}$  &Fe$^{3+}$-Fe$^{3+}$\\ \hline
HTP		&2.96			&\textendash\         & \textendash\ \\ 
LTP		&2.95                   &2.89/3.03	      &2.95 \\ \hline\hline
\end{tabular}

\end{center}
\end{table}
The inhomogeneities in the d$^{FF}$ at the LTP have important consequences for the emergence 
of trimerons. Regarding figure \ref{fbulk-trims},
every Fe$^{2+}$ is surrounded by 4 Fe$^{3+}$ placed at the adjacent upper and lower (001) layers.
Two of them are at 2.89$\text{\AA}$ (solid colored lines) and the other two are at 3.03$\text{\AA}$ (dotted lines),
while the interatomic distance between coplanar Fe$_B$ atoms is 2.95$\text{\AA}$, as shown in table \ref{tabla-dist}.
This defines linear Fe$^{3+}$-Fe$^{2+}$-Fe$^{3+}$ chains of shortened lengths, with a charge accumulation over 0.027 $e/\text{\AA}^3$
at the middle of each Fe$^{3+}$-Fe$^{2+}$ segment, in analogy with the experimental features assigned to trimerons
\cite{Attfield-nature}.
The orbital character of the electronic states confirms the polaronic charge distribution, with the occupied 
Fe$^{2+}$ $t_{2g}$ minority spin states lying along the central axis of the
chain and inducing a small contribution of the same orbital character at the closer Fe$^{3+}$.

Trimerons are uniquely characterized by the coexistence of all these features $-$short d$^{FF}$ ($<2.93$ $\text{\AA}$),
enhanced charge accumulation ($>0.027$ e/$\text{\AA}^3$) and orbital directionality$-$, as confirmed by exploring 
alternative solutions without CO along the (001) direction where trimerons do not form.
Moreover, the existence of these solutions points to the complex link between the long- and short-range COs \cite{Piekarz-2014}.
We have observed that,
already under a cubic lattice, the reduced $P2/m$ unit cell is enough for the Verwey metal-insulator transition to emerge, 
merely by relaxing the symmetry constraints of the HTP in the presence of electronic correlations 
(U$> 2$ eV) \cite{Leonov-2004}.
The additional full relaxation of the lattice vectors and atomic positions introduces a slight orthorhombic distortion
at the LTP, and is accompanied by the formation of the local trimeron structures.
This links the metal-insulator transition to the extended CO, and separates it from the short-range correlations, 
in good agreement with recent evidence \cite{Piekarz-2014}.
Though the intricate relation between the different COs can only be ultimately integrated under the full $Cc$ symmetry,
the results presented in this section prove that our reduced unit cell contains the main features of the charge distribution 
at the LTP: a dominant CO along the $[001]$ axis \cite{Wright-2002}, and the existence of trimerons as short-range
features that are distinct to the low temperature CO but intimately connected to it. 
This supports the use of the $P2/m$ cell as a basis to explore the surface properties below T$_V$.

\section{The $\text{Fe}_3\text{O}_4$(001) surface above T$_V$}

\begin{figure}[htbp]
\begin{center}
\includegraphics[width=0.75\columnwidth,clip]{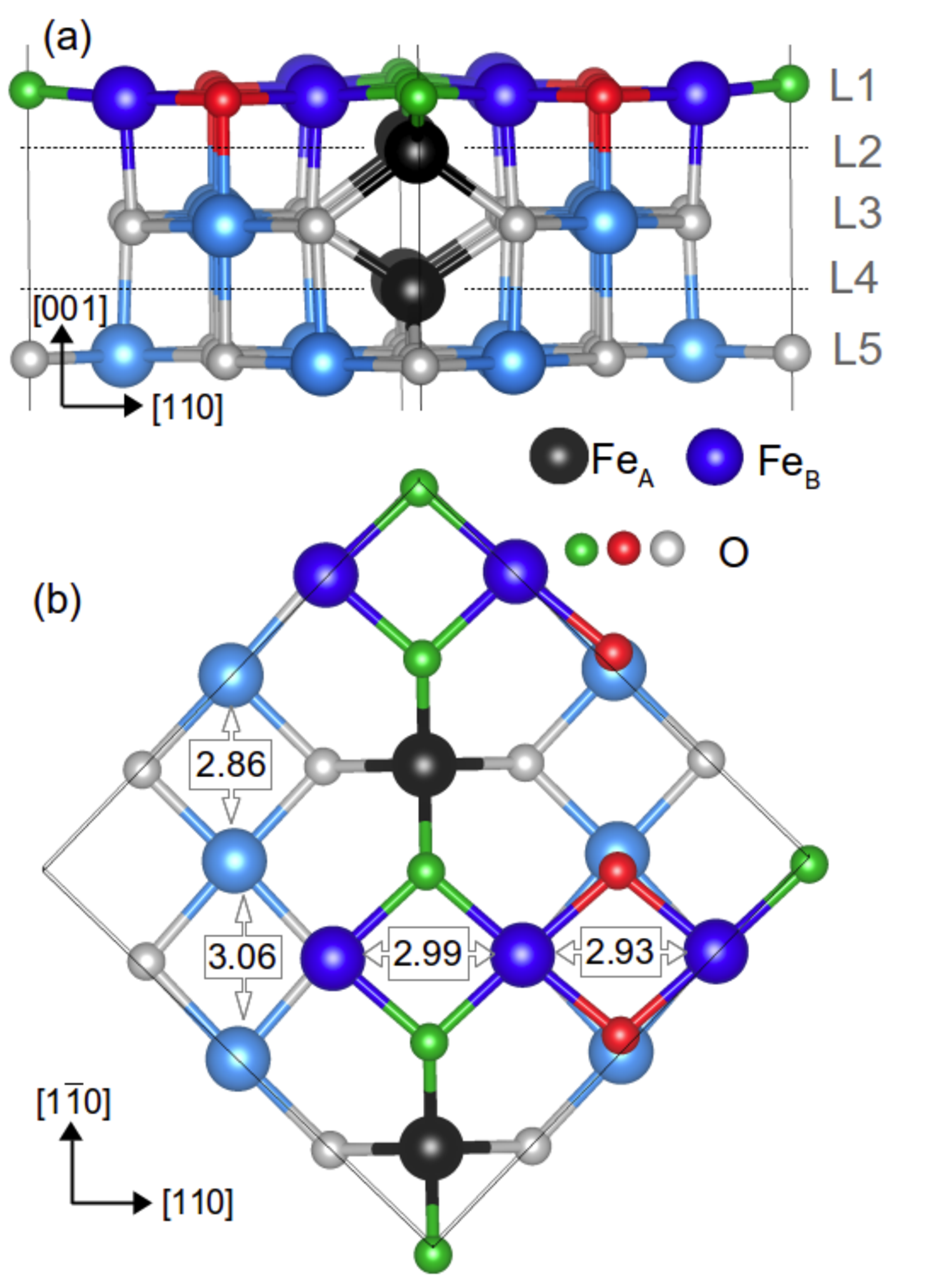}
\caption{
(Color online)
(a) Top and (b) side views of the Fe$_3$O$_4$(001) surface above T$_S$.
Panel (b) only shows the 3 outermost planes,
indicating the different in-plane Fe$_B$-Fe$_B$ distances in $\text{\AA}$.
Also, for clarity, the leftmost surface row of O atoms is not depicted.}
\label{fsurf-struc1}
\end{center}
\end{figure}

We will first focus on the unreconstructed surface of the HTP above T$_S$.
A sketch of the structure corresponding to our ground state is depicted in figure \ref{fsurf-struc1}, where 
layers are numbered from the surface (L1) towards the bulk.
As each surface O atom has lost one donor neighbor,
they reduce the bond lengths to the remaining Fe$_B$ cations to $\sim 1.97$ $\text{\AA}$ in order to recover the bulk-like charge.
This leads to a significant rearrangement of the atomic positions,
where the compression of the first interlayer distance (d$_{12} =0.78$ $\text{\AA}$,
to be compared to the bulk value $1.04$ $\text{\AA}$) is followed by the expansion of 
the subsequent interlayer spacings (d$_{23}=1.17$ $\text{\AA}$, d$_{34}=1.07$ $\text{\AA}$).
As indicated in figure \ref{fsurf-struc1}, 
at L1 there are two types of O sites, either bonded to a subsurface Fe$_B$ (O$_B$) or to Fe$_A$ (O$_A$).
In order to avoid the excessive shortening of the O-Fe$_A$ distance, 
the O$_A$ atoms move outwards, inducing at L1
a large corrugation of 0.11 $\text{\AA}$ 
and a slight in-plane wavelike distortion of the O rows.
The asymmetry persists at L3, where the O corrugation attenuates to 0.04 $\text{\AA}$.
While Fe$_A$ remains essentially unaffected by the large distortion of the O sublattice, 
the opposite occurs for the Fe$_B$ at the two outermost layers, L1 and L3. Each Fe$_B$ 
along the surface $[110]$ and subsurface $[1\overline{1}0]$ rows
approaches one of their adjacent Fe neighbors at the cost of farthening from the opposite.
As shown in figure \ref{fsurf-struc1}(b), the movement is more pronounced at the subsurface.

\begin{figure}[htbp]
\begin{center}
\includegraphics[width=0.8\columnwidth,clip]{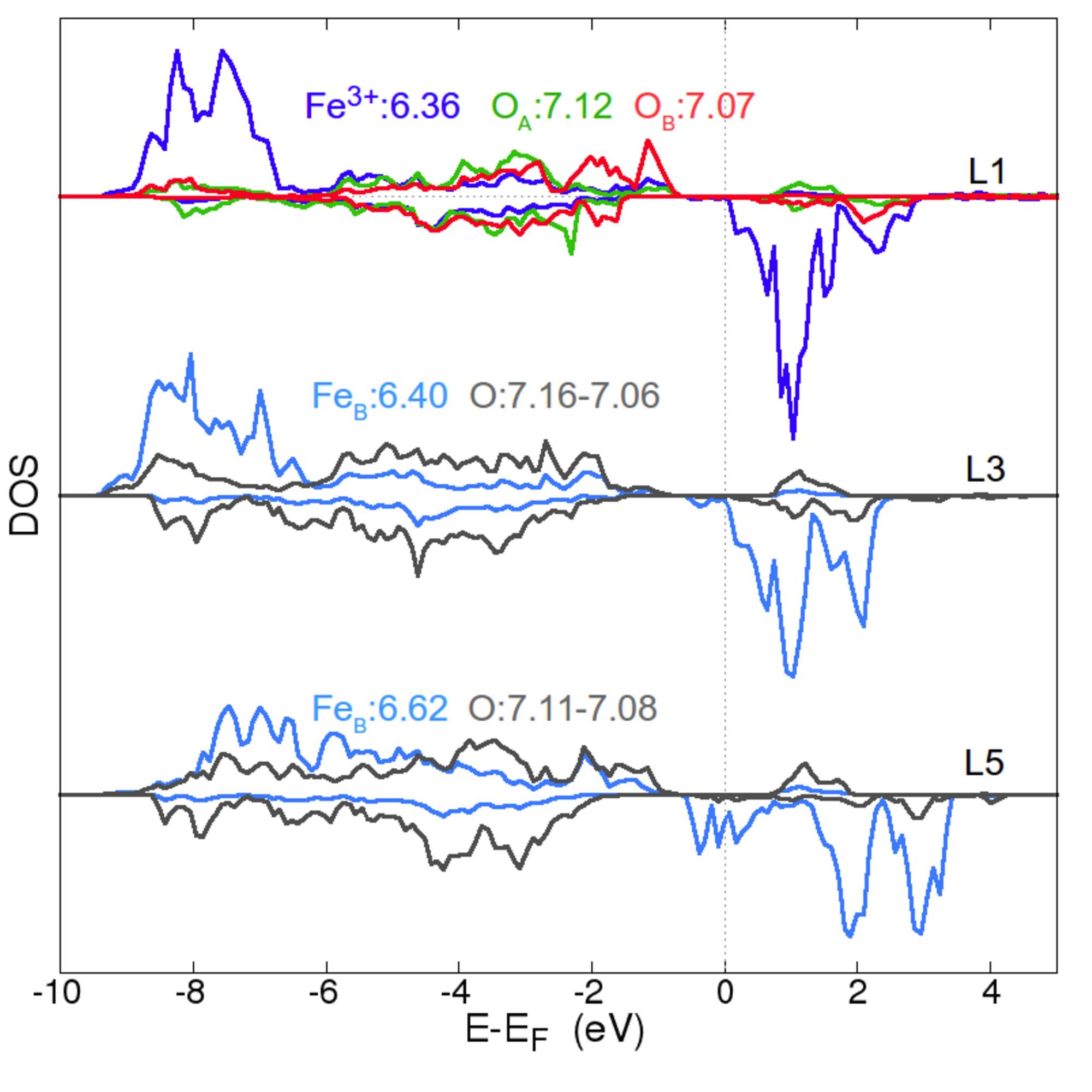}
\caption{
(Color online)
Spin-resolved DOS of all inequivalent atoms (blue for Fe, red/green for surface O) 
at the outermost Fe$_B$-O planes of figure \protect\ref{fsurf-struc1},
providing the corresponding Q$_B$.}
\label{fsurf-struc}
\end{center}
\end{figure}
The result of this restructuration in the electronic properties can be seen in figure \ref{fsurf-struc},
that provides the atomic-resolved DOS and the corresponding Q$_B$ at the outermost Fe$_B$-O planes, 
where all surface effects are contained.
Although the O charges show significant dispersion, the 
differences are not apparent in the DOS, and their Q$_B$ are close to bulk values throughout the structure.
The Fe$_B$ atoms at L1 behave as Fe$^{3+}$, opening an insulating gap. 
However, the emergence of the gap is not accompanied by any charge disproportionation at the Fe$_B$ sublattice.
A gradual recovery of bulk-like behavior starts at L3, and is almost restored at L5.
As our slabs are not completely free from confinement effects, we cannot discard that it could be restored even at L3,
as inferred from STM observations of the structure of antiphase boundaries (APB) \cite{Norm-2013,Parkinson-2012}.
It is also important to remark that although
uncompensated and slightly enhanced magnetic moments emerge at the surface plane (4.16$\mu_B$ for Fe, 0.4$\mu_B$ for O),
the antiferromagnetic coupling between the Fe$_A$ and Fe$_B$ sublattices 
remains unaltered.
This preserves the bulk-like high magnetic moment of Fe$_3$O$_4$ also at the high temperature surface,
validating it as a promising material for spintronics applications.

\begin{figure}[htbp]
\begin{center}
\includegraphics[width=0.75\columnwidth,clip]{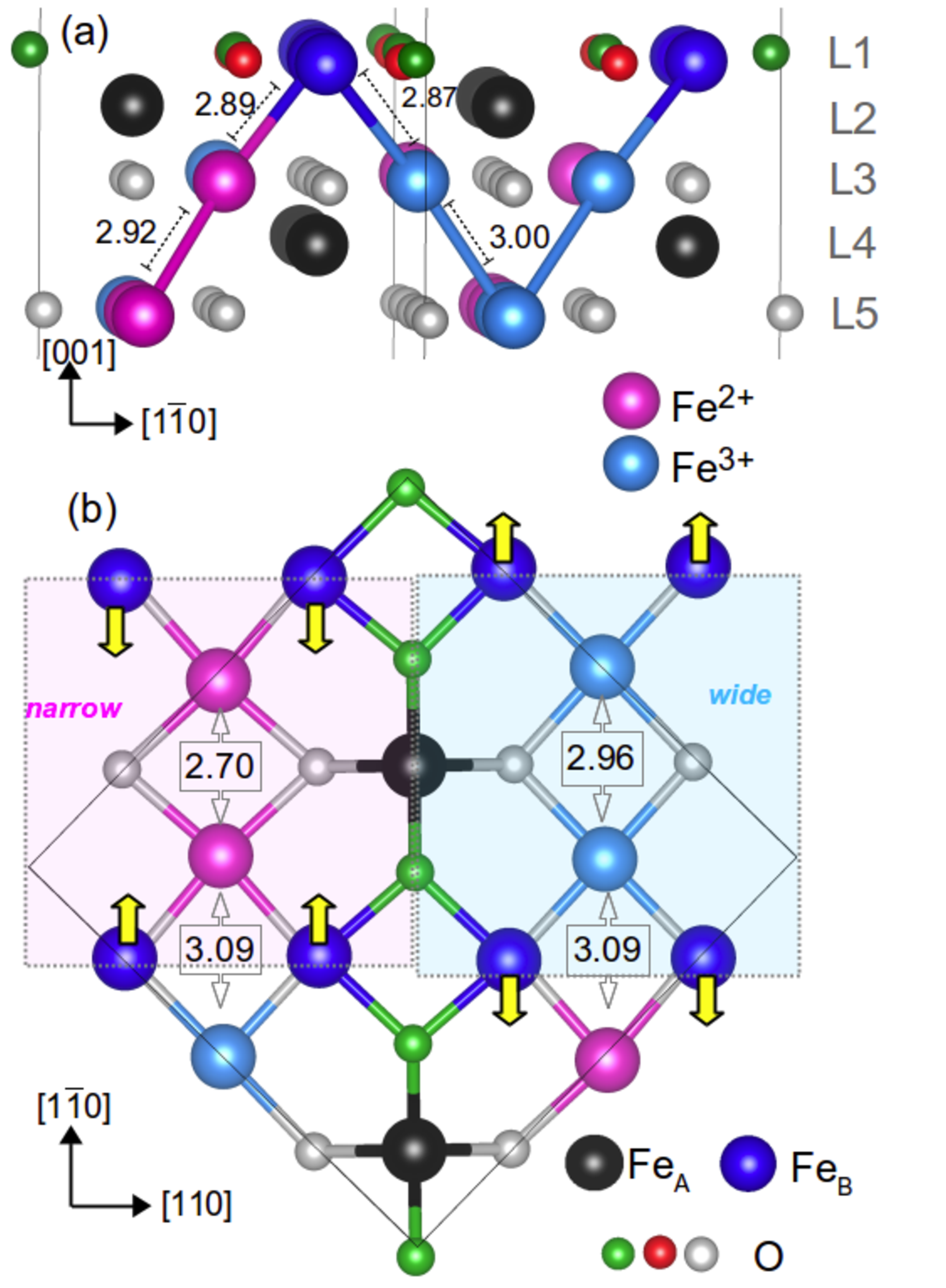}
\caption{
(Color online)
Same as figure \protect\ref{fsurf-struc1} for the HTP below T$_S$.
Values of the Fe$_B$-Fe$_B$ distances in $\text{\AA}$ are provided (a) between planes
and (b) along the surface and subsurface Fe$_B$ rows.
Arrows in (b) are a guide to indicate the wavelike displacements of surface Fe, and only
the central row of surface O atoms is depicted for clarity.}
\label{fsurf-is2-dos1}
\end{center}
\end{figure}
\begin{figure}[htbp]
\begin{center}
\includegraphics[width=0.8\columnwidth,clip]{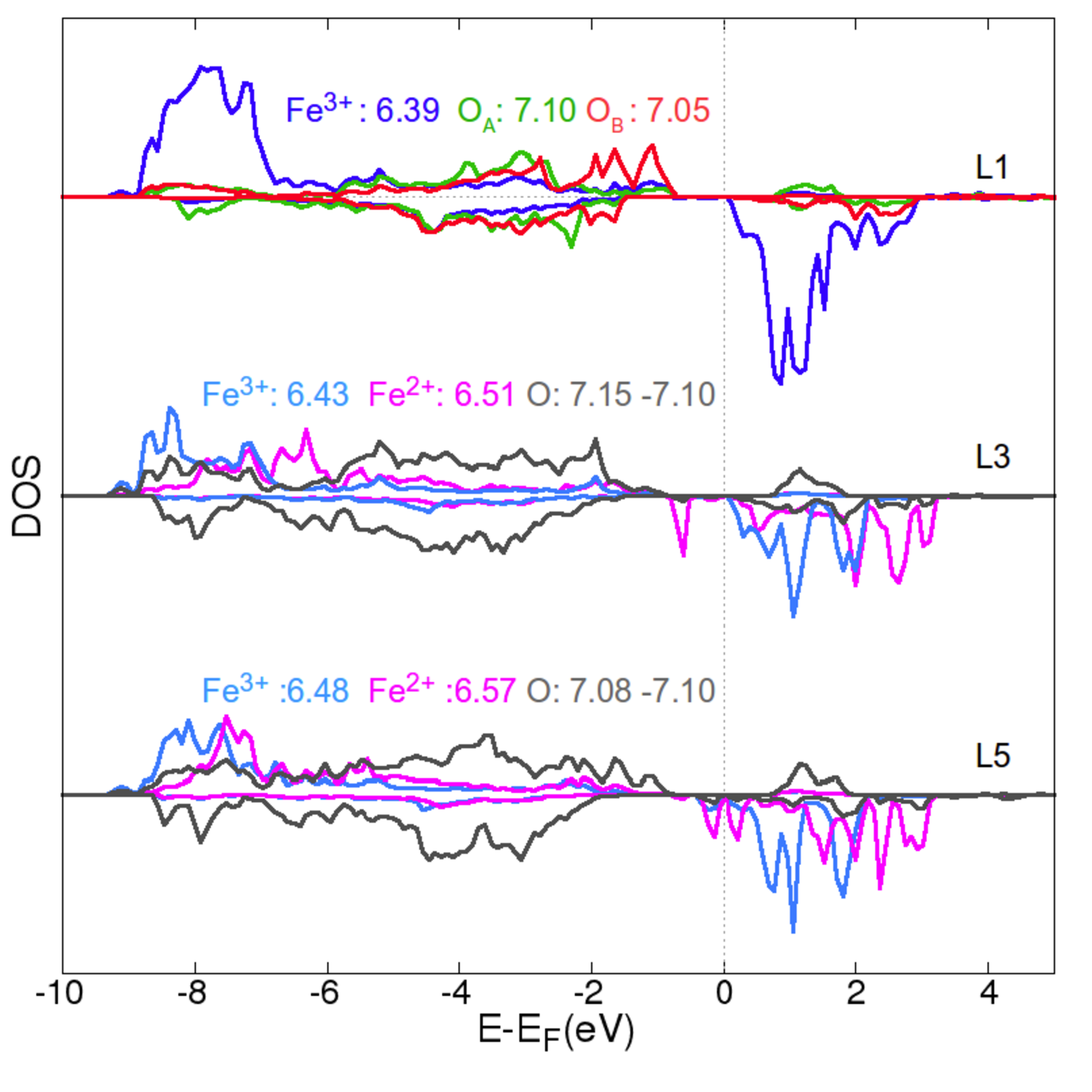}
\caption{
(Color online)
Same as figure \protect\ref{fsurf-struc} for the structure in figure \protect\ref{fsurf-is2-dos1}.} 
\label{fsurf-is2-dos}
\end{center}
\end{figure}

When the temperature is lowered below T$_S$, the $(\sqrt{2}\times\sqrt{2})R45^\circ$ reconstruction sets in.
This surface has already been studied in detail, but there are yet controversies about the origin of the reconstruction
and its dependence on electronic correlations \cite{Lodziana-2007,Pentcheva-2005}. Our results indicate that all
effects described for the unreconstructed surface are still present below T$_S$,
with only minor modificationis of d$_{23}$ and d$_{34}$ of less than 0.04 $\text{\AA}$, 
and slightly more
asymmetric Fe-O coordination units.
Figure \ref{fsurf-is2-dos1} shows a sketch of the structure 
and figure \ref{fsurf-is2-dos} the corresponding DOS and Q$_B$ at the 3 outermost
Fe$_B$-O planes.
The most relevant feature introduced by the reconstruction is the emergence of a charge disproportionation of 
$\sim 0.10$$e$ between Fe sites at L3, 
that defines a CO pattern within the (001) plane reducing the dispersion of O charges.
This subsurface CO was already proposed on the basis of purely electronic effects \cite{Lodziana-2007}.
However, we obtain that the
atomic wavelike displacement at the surface Fe rows lowers the energy by 28 meV/f.u. 
with respect to the $(1 \times 1)$ surface also in the presence of electronic correlations.
Reminiscence of this CO persists at L5, though half-metallicity is recovered.
In fact, we cannot discard some penetration of the surface effects at deeper layers in real samples,
where the existence of defects or APB may contribute to alterations of
the CO, as the energy barrier between different charge distributions is of only a few meV \cite{Parkinson-2012}.

From these results it is clear that the surface transition arises from the interplay between CO and electron-lattice
couplings, as already proposed on the basis of thermodynamic models \cite{Norm-2013}.
But although an insulating and charge-ordered state exists below T$_S$, the surface introduces significant differences
with the bulk LTP.
At L3, the Fe charge and DOS width are influenced by the demand of charge from surface O, and show
reduced values with respect to the bulk Fe$^{3+}$ and Fe$^{2+}$. More important, as we will prove now,
neither the surface structure nor the 
orbital character of the surface $t_{2g}$ states support the definition of trimerons.

Regarding figure \ref{fsurf-is2-dos1},
the wavelike Fe$_B$ surface displacements define narrow and wide regions occupied respectively by Fe$^{2+}$ and Fe$^{3+}$.
As a result, along each subsurface $[1\overline{1}0]$ row, pairs of Fe$^{2+}$ and Fe$^{3+}$ alternate, inhibiting
the formation of linear Fe$^{3+}$-Fe$^{2+}$-Fe$^{3+}$ chains within the (001) plane.
Eventhough the longitudinal movement of the surface Fe$_B$ along (110) rows (not shown in the figure) is similar to that above T$_S$, 
at the subsurface the
displacement of Fe$^{3+}$ is suppressed, 
originating shortened d$^{FF}$=2.70 $\text{\AA}$ between Fe$^{2+}$-Fe$^{2+}$ and large d$^{FF}$=3.09 $\text{\AA}$
between Fe$^{3+}$-Fe$^{2+}$.
This leads to in-plane charge sharing between Fe$^{2+}$ sites, forming a kind of localized bipolarons 
\cite{Lodziana-2007,Shvets-2004}
with a large charge accumulation of 0.035 $e/\text{\AA}^3$,
but opposes to the structure of bulk trimerons.
This tendency persists with respect to the adjacent planes: as shown in figure \ref{fsurf-is2-dos1}(a),
the d$^{FF}$ to the Fe$_B$ neighbors at L1
is similar for Fe$^{3+}$ and Fe$^{2+}$, and much larger than 2.70 $\text{\AA}$.
Similarly, the Fe$_B$ closer to subsurface Fe$^{3+}$ (Fe$^{2+}$) at L5 are those of Fe$^{3+}$ (Fe$^{2+}$) type, and are 
also farther than 2.70 $\text{\AA}$.
In conclusion, neither the interatomic distances nor the charge distribution arising from the surface
reconstruction support the formation of bulk-like trimerons.

\section{The $\text{Fe}_3\text{O}_4$(001) surface below T$_V$}

The different nature of the low temperature surface and bulk phases discards that the 
$(\sqrt{2}\times\sqrt{2})R45^o$ reconstruction acts as the first stage for the development of the VT. 
However, it suggests the possibility of a competition between surface and bulk COs
below T$_V$. In order to explore this, we have modelled the Fe$_3$O$_4$(001) surface of the LTP departing from
our $P2/m$ bulk unit cell. Although this cell contains limited information of the actual long-range CO,
we will show that yet important insights about the mutual influence of the bulk and surface short-range 
correlations become evident. 
\begin{figure}[htbp]
\begin{center}
\includegraphics[width=0.75\columnwidth,clip]{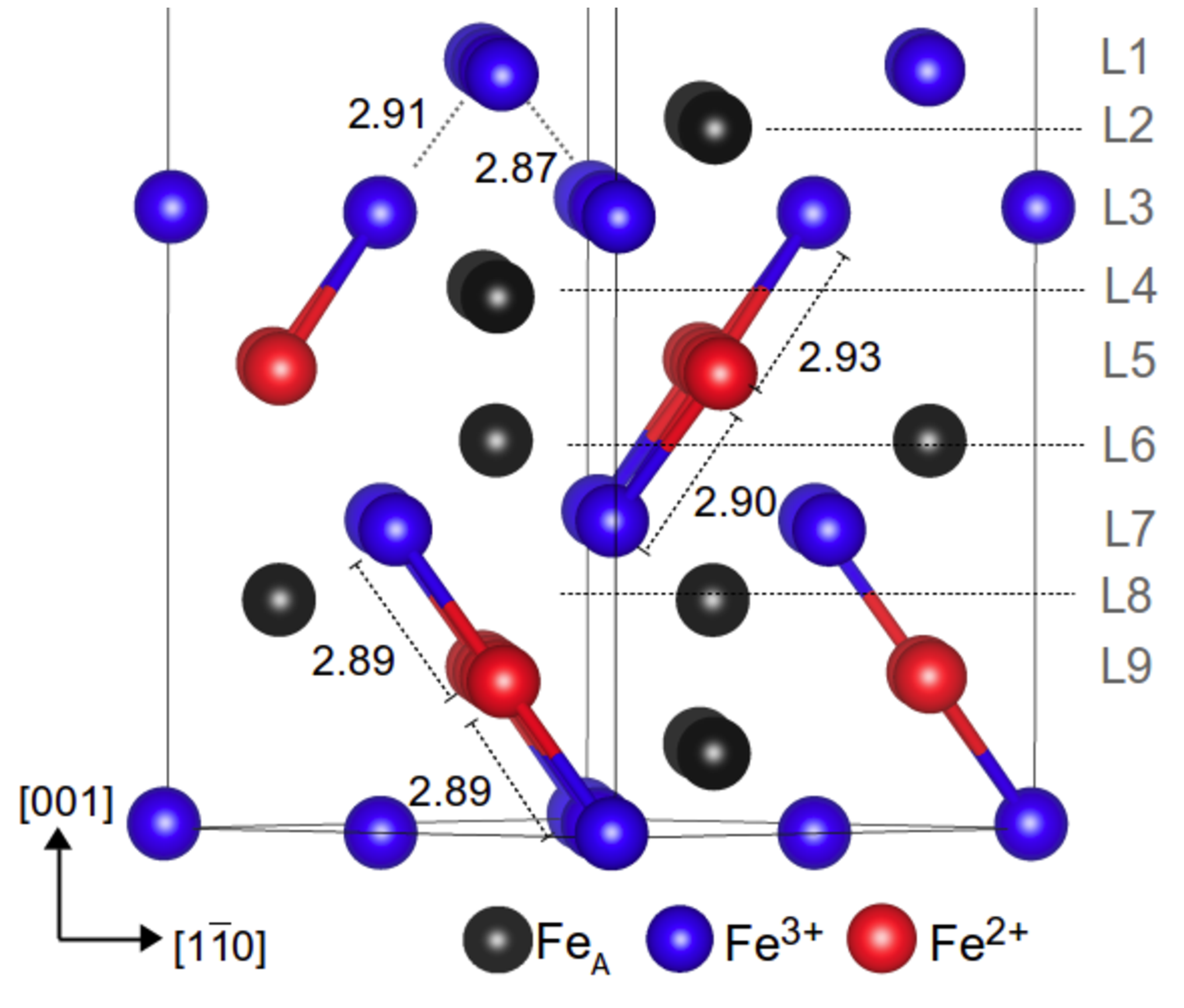}
\caption{
(Color online)
Side view of the Fe sublattice at the Fe$^{2+}$-ended Fe$_3$O$_4$(001) surface below T$_V$,
indicating the bulk-like trimerons and providing selected d$^{FF}$ values (in $\text{\AA}$).}
\label{fsurf-is01}
\end{center}
\end{figure}
\begin{figure}[htbp]
\begin{center}
\includegraphics[width=0.8\columnwidth,clip]{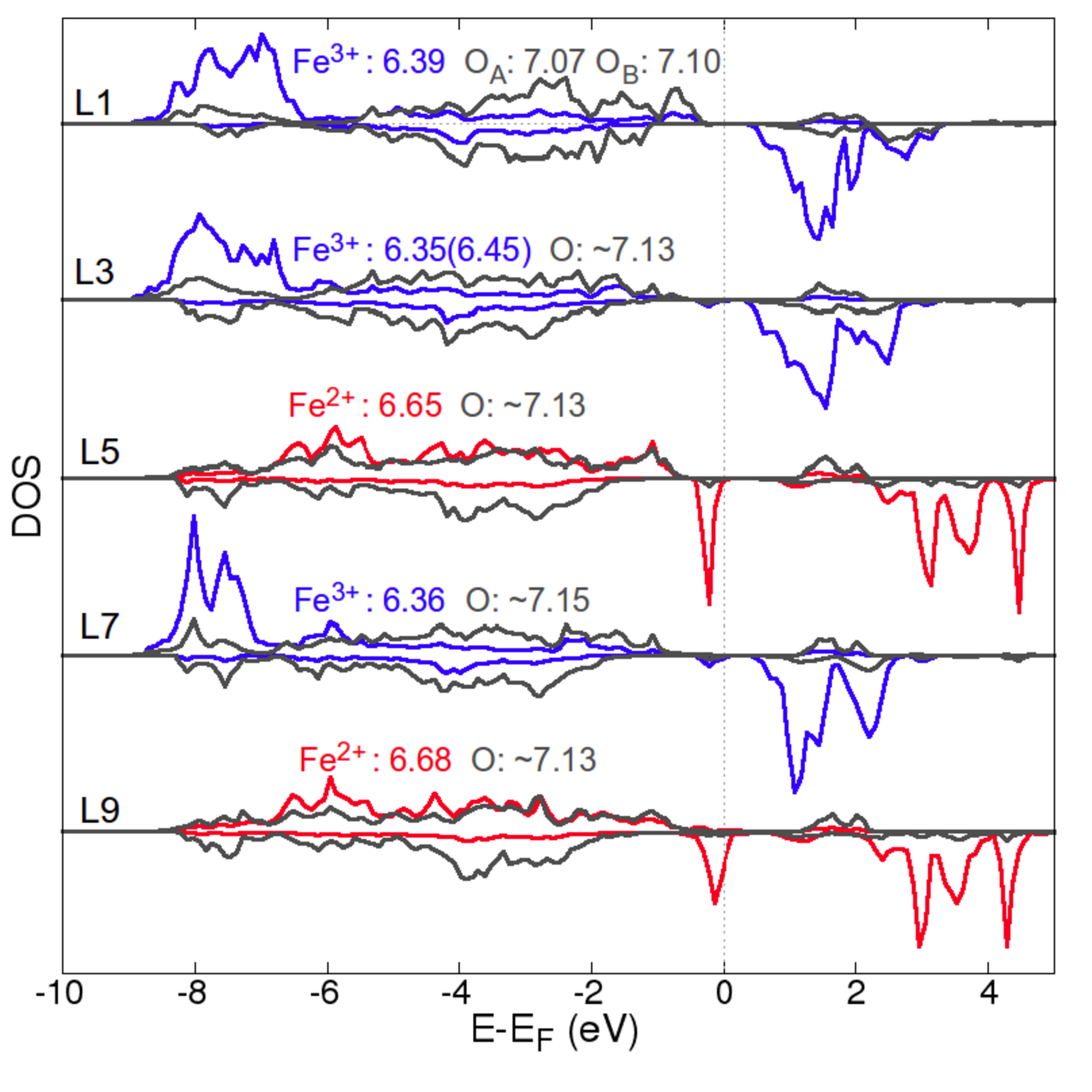}
\caption{
(Color online)
Layer- and spin-resolved DOS of the Fe$^{3+}$ (blue), Fe$^{2+}$ (red) and O (black) atoms of the structure 
in figure \ref{fsurf-is01}.}
\label{fsurf-is0}
\end{center}
\end{figure}
The surface can be constructed exposing either Fe$^{3+}$ or Fe$^{2+}$ planes, 
which has implications for the continuity of bulk trimerons close to the surface,
as shown in figures \ref{fsurf-is01} and \ref{fsurf-is0-Fe3+str}.
The most stable situation by $\sim 70$ meV/f.u. corresponds to the Fe$^{2+}$-ended case in figure \ref{fsurf-is01}, 
that at difference with the Fe$^{3+}$ termination, preserves the bulk CO up to the subsurface.
This energy difference is much larger than that between the
$(\sqrt{2}\times\sqrt{2})R45^o$ and $(1 \times 1)$ surfaces at high temperatures, evidencing the
high impact of the bulk CO on the surface properties.
We have estimated that the loss of the bulk CO at the subsurface lowers the work function by 0.30 eV, a variation close to
that induced by the adsorption of water \cite{Kendelewicz-2013}.

On the other hand, surface effects are similar to those at the HTP
under both terminations: an insulating Fe$^{3+}$ surface layer, shortened surface O-Fe$_B$
bonds, and a similar pattern of outermost interlayer distances and longitudinal atomic displacements within the Fe$_B$ rows.
This helps to attain bulk values of the O charge, though the surface causes an additional dispersion in Q$_B$,
as shown in figures \ref{fsurf-is0} and \ref{fsurf-is0-Fe3+} for the Fe$^{2+}$- and Fe$^{3+}$-ended cases, respectively.
Surprisingly, at the Fe$^{2+}$ termination the same electronic structure 
corresponds to the $(1 \times 1)$ and $(\sqrt{2}\times\sqrt{2})R45^o$
surfaces, separated by less than 7 meV/f.u.
This is because the LTP bulk structure introduces an additional charge modulation within (001) planes, 
that obscures that induced by the reconstruction: 
regarding figure \ref{fsurf-is01},
half of the Fe$^{3+}$ sites at L3 would develop trimerons with the upper Fe$_B$, but these have changed their valence inhibiting 
the polaronic charge distribution. As the other half participate in trimerons with the layers below, two types of Fe$_B$ sites exist 
at the subsurface, with similar DOS but slightly different Q$_B$ and interatomic distances to the surface Fe$_B$.
Again this proves the influence of the bulk CO on the surface properties below T$_V$.
\begin{figure}[htbp]
\begin{center}
\includegraphics[width=0.75\columnwidth,clip]{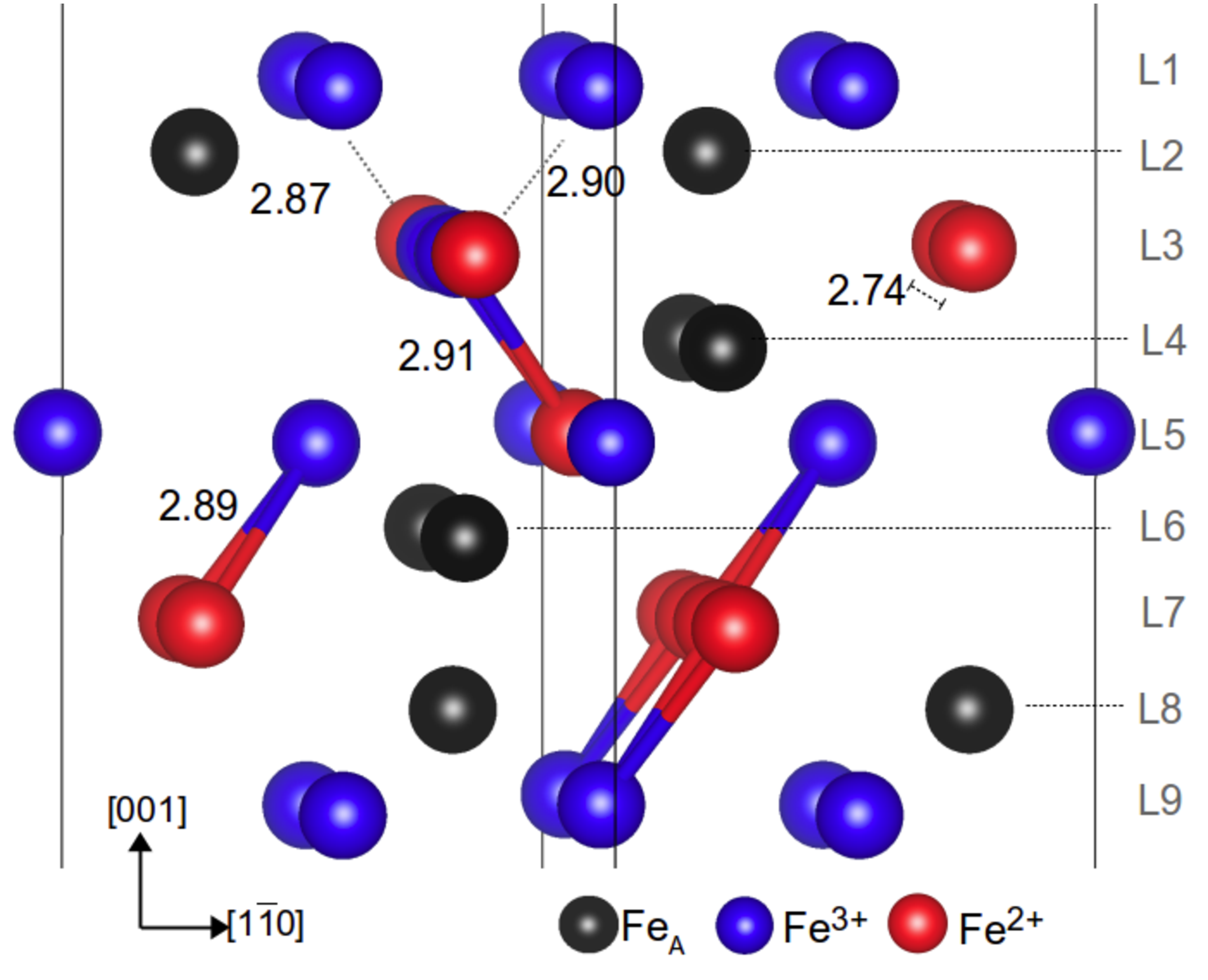}
\caption{
(Color online)
Same as figure \protect\ref{fsurf-is01} for the Fe$^{3+}$-ended surface of the LTP.}
\label{fsurf-is0-Fe3+str}
\end{center}
\end{figure}
\begin{figure}[htbp]
\begin{center}
\includegraphics[width=0.8\columnwidth,clip]{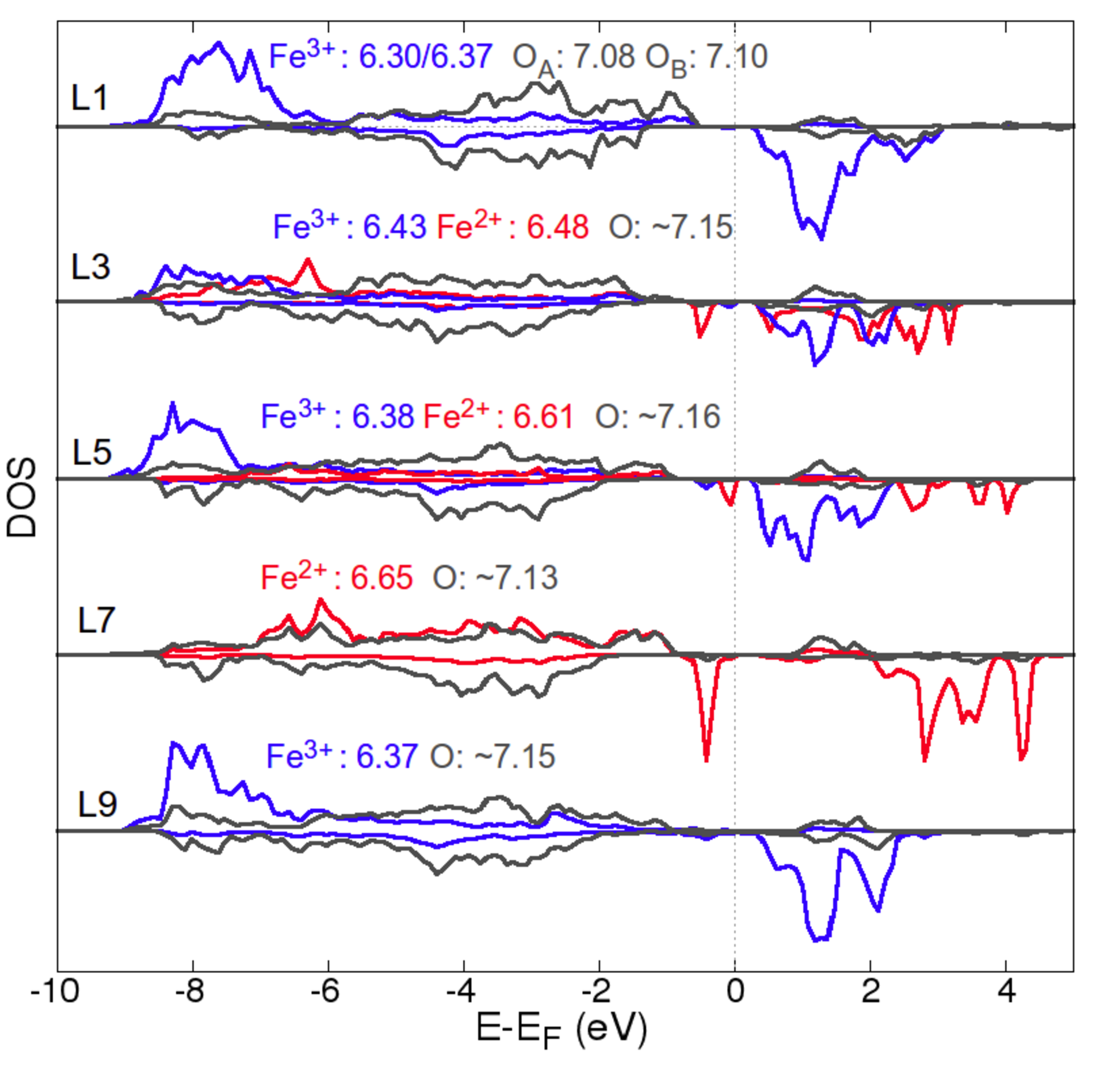}
\caption{
(Color online)
Same as figure \protect\ref{fsurf-is0} for the structure in figure \ref{fsurf-is0-Fe3+str}.}
\label{fsurf-is0-Fe3+}
\end{center}
\end{figure}
In turn, the robust insulating surface layer, which seems to be a universal feature of magnetite even under metastable
terminations \cite{Jordan-2006,Lodziana-2007}, has an important local effect on the bulk CO.
This better manifests at the Fe$^{3+}$ termination in figures \ref{fsurf-is0-Fe3+str} and \ref{fsurf-is0-Fe3+}, where the lack of continuity
of the trimerons at the subsurface allows for the emergence of localized bipolarons, indicating the possible coexistence
of local surface and bulk COs.
But figure \ref{fsurf-is01} evidences that also at the Fe$^{2+}$ termination those trimerons closer to the surface are slightly affected by it: the Fe$^{3+}$-Fe$^{2+}$ 
distances between L3 and L5 are moderately enlarged, which introduces an asymmetry in the
Fe chain weakening the charge sharing in its upper branch.
In summary, though preservation of the bulk CO seems to have a dominant effect on the surface stability,
it is conditioned by the insulating Fe$^{3+}$ surface layer, and there is a mutual influence of the bulk and
surface properties that extends several layers below the surface plane.

\section{Summary and conclusions}

Our results prove that the Fe$_3$O$_4$(001) surface shows a robust insulating state that persists across the surface and
bulk phase transitions.
It is originated by the large demand of charge from surface O arising from bond breaking,
and causes a significant restructuration at the outermost planes that inhibits the formation of trimerons.
Below T$_S$, a surface CO distinct from that of the bulk LTP emerges. Its distinct nature manifests 
in a lower charge disproportionation as compared to the bulk LTP, and in the preferential bipolaronic CO 
within (001) planes. When the temperature is lowered below T$_V$, this surface CO
competes with the dominant bulk one. This competition is conditioned by the 
insulating Fe$^{3+}$ character of the surface, which weakens the trimeron structures.

Besides its intrinsic interest, the relation between CO and dimensionality has implications for the multifunctional 
properties of magnetite, since
the emergence of ferroelectric polarization \cite{Yamauchi-2009}, the orientation of the magnetic easy axis \cite{JdF-roof} or the 
catalytic activity \cite{Skomurski-2010,Parkinson-catal-2013} 
are related to the existence of different Fe$_B$ sites and the resulting charge distribution.
Direct evidence from surface measurements is challenging, as most effects will manifest at the subsurface level.
Additional complications emerge from the existence of APB in real samples, and from differences in the relative orientation of 
the monoclinic and cubic crystal axes in single crystals and thin films.
From the theoretical side, the inclusion of the full $Cc$ symmetry, with additional modulations of the CO within (001) planes,
may show an even richer scenario.
However, our results unequivocally demonstrate the existence of a mutual influence of the surface and bulk COs, 
providing and partially quantifying the main features involved in it. We hope they will motivate further work in this fascinating subject.

\section*{Acknowledgments}

This work has been financed by the Spanish Ministry of Science under contracts MAT2009-14578-C03-03 and MAT2012-38045-C04-04. 
I.B. acknowledges financial support from the JAE program of the CSIC.

\providecommand{\noopsort}[1]{}\providecommand{\singleletter}[1]{#1}%

\end{document}